# Examining discontinuance of AI-mediated informal digital learning of English (AI-IDLE) among university students: Evidence from SEM and fsQCA

**First and Corresponding Author**
Yiran Du
University of Cambridge, Cambridge, UK
yd392@cam.ac.uk

**Second Author**
Huimin He
Xi'an Jiaotong-Liverpool University, Suzhou, China
Huimin.he@xjtlu.edu.cn

**Abstract**
This study examined university students' discontinuance intention towards AI-mediated informal digital learning of English (AI-IDLE). Drawing on the cognition–affect–conation framework, the study investigated how three cognitive factors, namely disconfirmation, perceived complexity, and perceived risk, influence two affective responses, namely dissatisfaction and frustration, and how these affective responses predict discontinuance intention. A cross-sectional survey was conducted with 746 Chinese university students who had experience using AI tools for informal English learning. Data were analysed using structural equation modelling (SEM) and fuzzy-set qualitative comparative analysis (fsQCA). The SEM results showed that dissatisfaction and frustration positively predicted discontinuance intention, with frustration showing the stronger effect. Disconfirmation, perceived complexity, and perceived risk also positively influenced dissatisfaction and frustration. The fsQCA results further identified multiple sufficient configurations leading to high AI-IDLE discontinuance intention, indicating that discontinuance is shaped by causal complexity and equifinality rather than by a single necessary condition. These findings extend AI-IDLE research from adoption and engagement to post-adoption disengagement and provide implications for reducing learners' dissatisfaction, frustration, perceived complexity, and risk in AI-supported informal English learning.

## 1. Introduction
AI-mediated informal digital learning of English (AI-IDLE) has become an important form of self-directed English learning beyond formal classrooms. Through generative AI tools, learners can obtain instant feedback, practise communication, revise texts, translate materials, and receive personalised language support, extending earlier forms of informal digital learning of English into a more interactive and adaptive learning ecology (Guan et al., 2025; Liu, Darvin, & Ma, 2025; Xia & Guo, 2025).

However, existing AI-IDLE research has mainly focused on adoption, engagement, and positive learning outcomes, while less attention has been paid to discontinuance after initial use (Liu, Ma, Bao, & Liu, 2025; Zadorozhnyy & Lee, 2025). This is an important gap because AI-IDLE is voluntary and weakly regulated by teachers or institutions. Learners may stop using AI tools when they experience unmet expectations, perceived complexity, perceived risk, dissatisfaction, or frustration (Li, 2025; Lin et al., 2020; Zhou & Wang, 2025; Zhou & Zhang, 2025).

To address this gap, this study examines university students' discontinuance intention towards AI-IDLE through the cognition–affect–conation framework. It investigates how disconfirmation, perceived complexity, and perceived risk influence dissatisfaction and frustration, and how these affective responses shape discontinuance intention. By combining structural equation modelling (SEM) and fuzzy-set qualitative comparative analysis (fsQCA), this study contributes to AI-IDLE and technology discontinuance research and offers practical implications for designing more sustainable, trustworthy, and pedagogically effective AI-supported English learning experiences.

## 2. Literature Review
### 2.1 AI-Mediated Informal Digital Learning of English (AI-IDLE)

Informal digital learning of English (IDLE) refers to learners' self-directed engagement with English through digital technologies outside formal instructional settings, such as through social media, online videos, games, translation tools, and communication platforms (Xia & Guo, 2025). Recent developments in generative artificial intelligence have extended this ecology into AI-mediated informal digital learning of English (AI-IDLE), where learners use tools such as ChatGPT, Bing Chat, DeepSeek, AI writing assistants, speech-recognition systems, and machine translation platforms to support English learning beyond the classroom (Zadorozhnyy & Lee, 2025). Compared with earlier forms of IDLE, AI-IDLE is more interactive, adaptive, and dialogic: learners can request explanations, receive immediate corrective feedback, practise writing or speaking, simulate conversations, translate texts, and personalise learning materials according to their own needs (Yang et al., 2025). This shift positions AI not merely as a digital resource but as a semi-autonomous learning mediator that can shape learners' access to input, opportunities for output, feedback quality, and perceived control over learning (Xia & Guo, 2025). Existing research has begun to conceptualise AI-IDLE as an emerging extension of IDLE, particularly in EFL contexts where learners may use AI tools to compensate for limited authentic English exposure and classroom interaction opportunities (Liu, Ma, et al., 2025).

However, because AI-IDLE occurs largely outside institutional regulation, its effectiveness depends heavily on learners' sustained voluntary engagement (Liu, Darvin, et al., 2025). This makes post-adoption processes especially important: even when learners initially adopt AI tools, their continued use may be disrupted by inaccurate responses, excessive cognitive effort, perceived over-reliance, privacy concerns, or a mismatch between expected and actual learning value (Guangxiang Leon Liu et al., 2025).

### 2.2 Discontinuance

Discontinuance refers to users' intention or behaviour to stop, suspend, reduce, or abandon the use of a technology after it has already been adopted (Zhou & Wang, 2025). In information systems research, discontinuance is conceptually distinct from non-adoption because it concerns the post-adoption stage, where users have accumulated direct experience and can evaluate whether the technology continues to meet their expectations (Zhou & Zhang, 2025). The expectation-confirmation tradition argues that users compare actual experience with prior expectations; when performance is confirmed, satisfaction and continuance are more likely, whereas negative disconfirmation can generate dissatisfaction and motivate discontinuance (Lin et al., 2020).

In AI-IDLE, discontinuance is particularly salient because informal English learning is voluntary, flexible, and weakly constrained by teachers, curricula, or assessment requirements (Guan et al., 2025). Learners can therefore disengage from AI tools quietly when the tools become frustrating, complex, risky, unreliable, or insufficiently useful for their learning goals (Y. Zhang & Liu, 2024). Discontinuance may be driven by affective responses, such as frustration or dissatisfaction, as well as cognitive appraisals, such as disconfirmation, perceived complexity, and perceived risk (Zhou & Wang, 2025). These mechanisms are especially relevant to AI-based language learning because learners must evaluate not only usability but also linguistic accuracy, pedagogical trustworthiness, data privacy, ethical acceptability, and the perceived authenticity of AI-mediated communication (Lee, Yeung, et al., 2024). Accordingly, examining AI-IDLE discontinuance can help explain why learners abandon technologies that initially appear useful, and it provides a necessary complement to adoption-oriented studies that mainly emphasise acceptance, motivation, and continuance.

## 3. Theoretical Framework and Hypothesis Development
### 3.1 Cognition–Affect–Conation Framework

The cognition–affect–conation (CAC) framework is adopted in this study because AI-IDLE discontinuance is not merely a technical decision but a post-adoption behavioural response shaped by learners' evaluations and emotional experiences. CAC is particularly suitable for modelling this process because it explains how individuals' cognitive appraisals of an experience are translated into affective responses and then into behavioural intentions (Zhou & Zhang, 2025). This logic is consistent with the present study's focus on why learners who have already used AI tools for informal English learning may intend to reduce or stop their use. As shown in Figure 1, the conceptual model positions disconfirmation, perceived complexity, and perceived risk as cognitive factors; dissatisfaction and frustration as affective factors; and discontinuance intention as the conative outcome. The CAC framework therefore provides a coherent theoretical structure for examining the pathways through which negative post-adoption evaluations and affective responses contribute to AI-IDLE discontinuance.

**Figure 1. The Conceptual Model**
**Alt text:** Conceptual model based on the cognition–affect–conation framework. Disconfirmation, perceived complexity, and perceived risk are shown as cognitive factors predicting dissatisfaction and frustration. Dissatisfaction and frustration are shown as affective factors predicting learners' discontinuance intention towards AI-mediated informal digital learning of English.

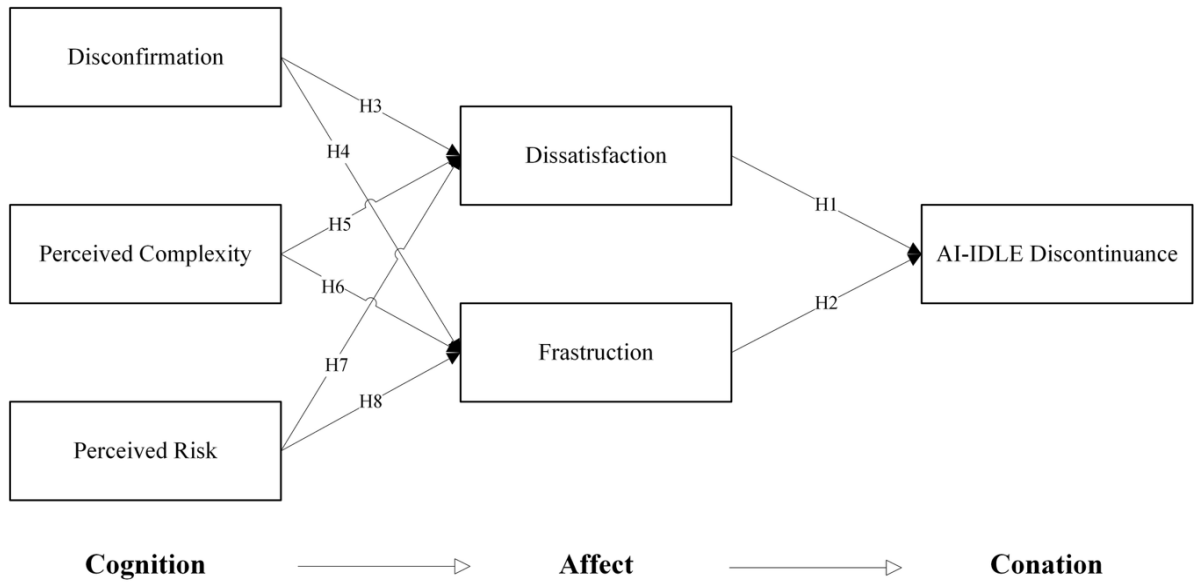

### 3.2 The Impact of Dissatisfaction and Frustration

Dissatisfaction in this context refers to learners' negative overall evaluation of their AI-IDLE experience after comparing actual use with their expectations or learning needs (Liu, Ma, et al., 2025; T. Zhang & Tong, 2025). In post-adoption contexts, dissatisfaction is a critical affective response because it signals that continued use is no longer perceived as worthwhile, rewarding, or beneficial (Song & Zhou, 2026). In AI-IDLE, learners may become dissatisfied when AI tools fail to provide accurate explanations, useful feedback, meaningful interaction, or sufficient support for informal English learning (Lee, Xie, et al., 2024). As AI-IDLE is largely voluntary and self-directed, dissatisfaction may directly weaken learners' motivation to continue using AI tools and increase their intention to discontinue use (R. Wu, 2023).

Frustration in this context refers to learners' negative emotional reaction when they experience obstacles, interruptions, or difficulties during AI-IDLE (Novak et al., 2022; Xia & Guo, 2025). Such frustration may emerge from inaccurate AI-generated content, irrelevant responses, repeated prompting, technical instability, limited transparency, or difficulty judging the reliability of AI output (Novak et al., 2022). Compared with dissatisfaction, which reflects a broader evaluative judgement, frustration captures a more immediate emotional response to impeded learning or unsuccessful interaction (Novak et al., 2023). When learners repeatedly feel frustrated during AI-IDLE, they may perceive the activity as burdensome rather than supportive, thereby becoming more likely to reduce, suspend, or abandon their use of AI tools for informal English learning (Soyoof et al., 2023). Based on this reasoning, the following hypotheses are proposed:

H1: Dissatisfaction positively influences learners' discontinuance intention towards AI-IDLE.
H2: Frustration positively influences learners' discontinuance intention towards AI-IDLE.

### 3.3 The Impact of Disconfirmation, Perceived Complexity, and Perceived Risk

Disconfirmation in this context refers to the extent to which learners' actual AI-IDLE experiences fall short of their prior expectations (Marikyan et al., 2023; Zadorozhnyy & Lee, 2025). In post-adoption contexts, negative disconfirmation is likely to generate adverse affective responses because users recognise a gap between what they expected a technology to provide and what it actually delivers (Carraher-Wolverton & Hirschheim, 2023). In AI-IDLE, learners may expect AI tools to offer accurate explanations, personalised feedback, natural interaction, and efficient English learning support (Ranjan & Mukherjee, 2026). When these expectations are not met, they may become dissatisfied with the overall learning experience and frustrated by the discrepancy between anticipated and actual performance (Soyoof et al., 2023).

Perceived complexity in this context refers to learners' perception that AI-IDLE is difficult to understand, manage, or use effectively (Hmoud et al., 2023; Y. Zhang & Liu, 2025). Although AI tools are often presented as accessible learning resources, learners may still experience complexity when they need to design effective prompts, interpret AI-generated responses, verify linguistic accuracy, or integrate AI support into their own informal learning routines (Guan et al., 2025). Such complexity can reduce the perceived ease and value of AI-IDLE, thereby increasing dissatisfaction and frustration (Liu, Darvin, et al., 2025).

Perceived risk in this context refers to learners' concerns about the potential negative consequences of using AI tools for informal English learning, including risks related to inaccurate output, privacy, over-reliance, academic integrity, and reduced learning autonomy (X. Wang et al., 2025; W. Wu et al., 2022). When learners perceive these risks as salient, they may feel less confident in AI-IDLE as a trustworthy learning approach, which can intensify dissatisfaction and frustration (Li, 2025). Accordingly, disconfirmation, perceived complexity, and perceived risk are expected to function as cognitive antecedents of learners' negative affective responses in AI-IDLE. Based on this reasoning, the following hypotheses are proposed:

H3: Disconfirmation positively influences learners' dissatisfaction with AI-IDLE.
H4: Disconfirmation positively influences learners' frustration with AI-IDLE.
H5: Perceived complexity positively influences learners' dissatisfaction with AI-IDLE.
H6: Perceived complexity positively influences learners' frustration with AI-IDLE.
H7: Perceived risk positively influences learners' dissatisfaction with AI-IDLE.
H8: Perceived risk positively influences learners' frustration with AI-IDLE.

### 4. Methodology
### 4.1 Research Design

This study adopted a quantitative research design to examine the discontinuance of AI-mediated informal digital learning of English (AI-IDLE) among university students. A cross-sectional survey was used to collect data from learners who had experience using AI tools for informal English learning outside formal classroom settings. The questionnaire measured learners' cognitive evaluations, affective responses, and discontinuance intention based on the proposed cognition–affect–conation framework. To provide both symmetric and configurational evidence, the study employed a dual analytical approach combining structural equation modelling (SEM) and fuzzy-set qualitative comparative analysis (fsQCA). SEM was used to test the net effects among the proposed constructs and evaluate the hypothesised relationships in the conceptual model, whereas fsQCA was used to identify configurations of conditions that may jointly lead to high AI-IDLE discontinuance intention. This design enabled the study to examine not only whether individual factors significantly predict discontinuance intention, but also how different combinations of cognitive and affective conditions contribute to learners' discontinuance of AI-IDLE.

### 4.2 Participants

Participants were recruited through Credamo, a professional online survey platform in China. Eligible participants had to be university students with prior experience using AI tools for informal English learning outside formal classroom requirements. A total of 812 responses were initially collected, of which 66 were excluded because respondents did not meet the eligibility criteria, failed attention-check items, completed the survey too quickly, provided patterned responses, or submitted incomplete questionnaires (Ward & Meade, 2023). The final sample comprised 746 valid participants. As shown in Table 1, the sample included 313 males and 433 females; 298 participants were aged 18–20, 337 were aged 21–23, and 111 were aged 24 or above. Most participants were undergraduates (*n* = 582), while 164 were postgraduates; 343 were from STEM disciplines and 403 from non-STEM disciplines. Before completing the survey, all participants provided informed consent and were informed that participation was voluntary, anonymous, and confidential.

**Table 1. Participant Characteristics (*N* = 746)**

| Characteristic | Category | n | % |
| --- | --- | --- | --- |
| Gender | Male | 313 | 42.0 |
| | Female | 433 | 58.0 |
| Age | 18–20 | 298 | 39.9 |
| | 21–23 | 337 | 45.2 |
| | 24 or above | 111 | 14.9 |
| Study level | Undergraduate | 582 | 78.0 |
| | Postgraduate | 164 | 22.0 |
| Academic discipline | STEM | 343 | 46.0 |
| | Non-STEM | 403 | 54.0 |

### 4.3 Measurement

The questionnaire consisted of two sections. The first section collected demographic information, including gender, age, study level, and academic discipline. The second section measured the six latent constructs in the conceptual model: disconfirmation (Marikyan et al., 2023), perceived complexity (Hmoud et al., 2023), perceived risk (Li, 2025), dissatisfaction (T. Zhang & Tong, 2025), frustration (Novak et al., 2022), and discontinuance intention (Zhou & Wang, 2025). Each construct was measured using three items adapted from prior technology continuance and discontinuance research and modified for the context of AI-mediated informal digital learning of English. All items were rated on a five-point Likert scale ranging from 1 = strongly disagree to 5 = strongly agree, with higher scores indicating higher levels of the corresponding construct. The English items were translated into Chinese and checked through back-translation to ensure semantic equivalence (Klotz et al., 2023). To assess common method variance (CMV) (Podsakoff et al., 2024), Harman's single-factor test was conducted. The first unrotated factor explained 32.46% of the total variance, below the commonly used threshold of 50%, suggesting that CMV was unlikely to be a serious concern. The full list of constructs and measurement items is presented in Table 2.

**Table 2. Constructs and Measurement Items**

| Construct | Item | Measurement Item (English) | Measurement Item (Chinese) |
| --- | --- | --- | --- |
| Disconfirmation (Marikyan et al., 2023) | DIS1 | My experience of using AI tools for informal English learning was worse than I had expected. | 我使用 AI 工具进行非正式英语学习的体验比我原先预期的更差。 |
| | DIS2 | AI tools for informal English learning performed below my prior expectations. | AI 工具在非正式英语学习中的表现低于我原先的预期。 |
| | DIS3 | The learning support provided by AI tools was less effective than I had expected. | AI 工具提供的学习支持不如我原先预期的有效。 |
| Perceived Complexity (Hmoud et al., 2023) | PC1 | Learning to use AI tools effectively for informal English learning is difficult for me. | 学会有效使用 AI 工具进行非正式英语学习对我来说是困难的。 |
| | PC2 | Using AI tools for informal English learning requires too much effort. | 使用 AI 工具进行非正式英语学习需要投入过多精力。 |
| | PC3 | I find it difficult to manage my use of AI tools during informal English learning. | 在非正式英语学习中，我觉得很难管理自己对 AI 工具的使用。 |
| Perceived Risk (Li, 2025) | PR1 | I am concerned that AI tools may provide inaccurate English learning support. | 我担心 AI 工具可能提供不准确的英语学习支持。 |
| | PR2 | I am concerned that using AI tools may have negative consequences for my English learning. | 我担心使用 AI 工具可能会对我的英语学习产生负面影响。 |
| | PR3 | I am concerned about privacy or data security risks when using AI tools for informal English learning. | 使用 AI 工具进行非正式英语学习时，我担心隐私或数据安全风险。 |
| Dissatisfaction (T. Zhang & Tong, 2025) | DS1 | I am dissatisfied with my overall experience of using AI tools for informal English learning. | 我对使用 AI 工具进行非正式英语学习的整体体验感到不满意。 |
| | DS2 | My experience of using AI tools for informal English learning has been disappointing. | 我使用 AI 工具进行非正式英语学习的体验令人失望。 |
| | DS3 | I feel frustrated when AI tools fail to support my informal English learning effectively. | 当 AI 工具不能有效支持我的非正式英语学习时，我会感到沮丧。 |
| Frustration (Novak et al., 2022) | FR1 | I feel discouraged when using AI tools for informal English learning does not go smoothly. | 当使用 AI 工具进行非正式英语学习不顺利时，我会感到受挫。 |
| | FR2 | I feel emotionally strained when AI tools do not respond to my English learning needs. | 当 AI 工具不能回应我的英语学习需求时，我会感到情绪上的压力。 |
| | FR3 | I feel frustrated when AI tools fail to support my informal English learning effectively. | 当 AI 工具不能有效支持我的非正式英语学习时，我会感到沮丧。 |
| Discontinuance Intention (Zhou & Wang, 2025) | DI1 | I intend to stop using AI tools for informal English learning. | 我打算停止使用 AI 工具进行非正式英语学习。 |
| | DI2 | I intend to reduce my use of AI tools for informal English learning. | 我打算减少使用 AI 工具进行非正式英语学习。 |
| | DI3 | I intend to avoid using AI tools for informal English learning in the future. | 我打算在未来避免使用 AI 工具进行非正式英语学习。 |

### 4.4 Data Analysis

Data analysis was conducted in two stages. First, structural equation modelling (SEM) (Kline, 2023) was used to assess the measurement model and test the hypothesised relationships in the conceptual model. Descriptive statistics, normality, reliability, convergent validity, and discriminant validity were examined before testing the structural paths. Model fit was evaluated using $\chi^2/df$, CFI, TLI, RMSEA, and SRMR. Bootstrapping was also conducted to examine the mediating effects of dissatisfaction and frustration. Second, fuzzy-set qualitative comparative analysis (fsQCA) (Schneider & Wagemann, 2012) was performed to identify configurations of cognitive and affective conditions associated with AI-IDLE discontinuance. The variables were calibrated into fuzzy sets, followed by necessity analysis and truth-table analysis. This combined analytical strategy enabled the study to examine both the net effects of individual predictors and the configurational pathways leading to AI-IDLE discontinuance.

## 5. Results
### 5.1 Structural Equation Modelling (SEM)

Structural equation modelling (SEM) was conducted to test the hypothesised relationships in the conceptual model. The descriptive statistics are reported in Table 3. The six constructs showed moderate mean scores, with perceived risk having the highest mean ($M = 3.32$, $SD = 0.79$) and discontinuance intention having the lowest mean ($M = 2.87$, $SD = 0.91$). Skewness values ranged from −0.21 to 0.22, and kurtosis values ranged from −0.51 to −0.27, suggesting no serious deviation from normality. As shown in Table 4, both the measurement model and structural model demonstrated satisfactory fit. The measurement model showed good fit, $\chi^2/df = 2.14$, CFI = 0.961, TLI = 0.952, RMSEA = 0.039, and SRMR = 0.034. The structural model also showed acceptable fit, $\chi^2/df = 2.27$, CFI = 0.954, TLI = 0.946, RMSEA = 0.041, and SRMR = 0.038.

**Table 3. Descriptive Statistics of the Constructs**

| Construct | *M* | *SD* | Skewness | Kurtosis |
|---|---|---|---|---|
| Disconfirmation | 3.18 | 0.86 | -0.12 | -0.48 |
| Perceived complexity | 3.05 | 0.82 | 0.08 | -0.36 |
| Perceived risk | 3.32 | 0.79 | -0.21 | -0.27 |
| Dissatisfaction | 2.94 | 0.88 | 0.16 | -0.42 |
| Frustration | 3.01 | 0.84 | 0.10 | -0.39 |
| Discontinuance intention | 2.87 | 0.91 | 0.22 | -0.51 |

**Table 4. Model Fit Indices**

| Fit Index | Threshold | Measurement Model | Structural Model |
|---|---|---|---|
| $\chi^2/df$ | < 3.00 | 2.14 | 2.27 |
| CFI | > 0.90 | 0.961 | 0.954 |
| TLI | > 0.90 | 0.952 | 0.946 |
| RMSEA | < 0.08 | 0.039 | 0.041 |
| SRMR | < 0.08 | 0.034 | 0.038 |

Note. CFI = comparative fit index; TLI = Tucker–Lewis index; RMSEA = root mean square error of approximation; SRMR = standardised root mean square residual.

The reliability and convergent validity results are presented in Table 5. All standardised factor loadings exceeded 0.77, Cronbach's α values ranged from 0.83 to 0.88, composite reliability values ranged from 0.83 to 0.88, and AVE values ranged from 0.62 to 0.71, indicating acceptable reliability and convergent validity. Discriminant validity was also supported. As shown in Table 6, the square root of AVE for each construct was greater than its correlations with other constructs, satisfying the Fornell–Larcker criterion. In addition, the heterotrait–monotrait ratios reported in Table 7 were all below 0.85, further supporting discriminant validity.

**Table 5. Reliability and Convergent Validity**

| Construct | Item | Loading | Cronbach's α | CR | AVE |
|---|---|---|---|---|---|
| Disconfirmation | DIS1 | 0.84 | 0.86 | 0.86 | 0.67 |
| | DIS2 | 0.82 | | | |
| | DIS3 | 0.79 | | | |
| Perceived complexity | PC1 | 0.81 | 0.84 | 0.84 | 0.64 |
| | PC2 | 0.78 | | | |
| | PC3 | 0.80 | | | |
| Perceived risk | PR1 | 0.80 | 0.83 | 0.83 | 0.62 |
| | PR2 | 0.77 | | | |
| | PR3 | 0.79 | | | |
| Dissatisfaction | DS1 | 0.85 | 0.87 | 0.87 | 0.69 |
| | DS2 | 0.83 | | | |
| | DS3 | 0.81 | | | |

| | | | | | | |
|---|---|---|---|---|---|---|
| Frustration | | FR1 | 0.82 | 0.85 | 0.85 | 0.65 |
| | | FR2 | 0.80 | | | |
| | | FR3 | 0.79 | | | |
| Discontinuance intention | | DI1 | 0.86 | 0.88 | 0.88 | 0.71 |
| | | DI2 | 0.84 | | | |
| | | DI3 | 0.82 | | | |

Note. CR = composite reliability; AVE = average variance extracted.

**Table 6. Discriminant Validity (Fornell–Larcker Criterion)**

| Construct | DIS | PC | PR | DS | FR | DI |
|---|---|---|---|---|---|---|
| DIS | 0.82 | | | | | |
| PC | 0.42 | 0.80 | | | | |
| PR | 0.38 | 0.45 | 0.79 | | | |
| DS | 0.53 | 0.47 | 0.44 | 0.83 | | |
| FR | 0.49 | 0.51 | 0.46 | 0.58 | 0.81 | |
| DI | 0.41 | 0.43 | 0.39 | 0.55 | 0.57 | 0.84 |

Note. DIS = disconfirmation; PC = perceived complexity; PR = perceived risk; DS = dissatisfaction; FR = frustration; DI = discontinuance intention.

**Table 7. Discriminant Validity (Heterotrait–Monotrait Ratio)**

| Construct | DIS | PC | PR | DS | FR | DI |
|---|---|---|---|---|---|---|
| DIS | — | | | | | |
| PC | 0.51 | — | | | | |
| PR | 0.46 | 0.54 | — | | | |
| DS | 0.63 | 0.56 | 0.52 | — | | |
| FR | 0.59 | 0.61 | 0.55 | 0.69 | — | |
| DI | 0.49 | 0.52 | 0.47 | 0.65 | 0.67 | — |

Note. DIS = disconfirmation; PC = perceived complexity; PR = perceived risk; DS = dissatisfaction; FR = frustration; DI = discontinuance intention.

The structural model results are summarised in Table 8 and Figure 2. All eight hypotheses were supported. Dissatisfaction positively influenced discontinuance intention ($\beta = 0.31$, $SE = 0.05$, $z = 6.20$, $p < .001$), supporting H1. Frustration also positively influenced discontinuance intention ($\beta = 0.36$, $SE = 0.05$, $z = 7.20$, $p < .001$), supporting H2. Disconfirmation positively influenced dissatisfaction ($\beta = 0.34$, $SE = 0.04$, $z = 8.50$, $p < .001$) and frustration ($\beta = 0.22$, $SE = 0.05$, $z = 4.40$, $p < .001$), supporting H3 and H4. Perceived complexity positively influenced dissatisfaction ($\beta = 0.29$, $SE = 0.04$, $z = 7.25$, $p < .001$) and frustration ($\beta = 0.33$, $SE = 0.04$, $z = 8.25$, $p < .001$), supporting H5 and H6. Perceived risk had weaker but significant positive effects on dissatisfaction ($\beta = 0.18$, $SE = 0.06$, $z = 3.00$, $p < .01$) and frustration ($\beta = 0.16$, $SE = 0.06$, $z = 2.67$, $p < .01$), supporting H7 and H8.

**Table 8. Structural Model Results**

| Hypothesis | Path | $\beta$ | SE | z | Result |
|---|---|---|---|---|---|
| H1 | DS → DI | 0.31*** | 0.05 | 6.20 | Supported |
| H2 | FR → DI | 0.36*** | 0.05 | 7.20 | Supported |
| H3 | DIS → DS | 0.34*** | 0.04 | 8.50 | Supported |
| H4 | DIS → FR | 0.22*** | 0.05 | 4.40 | Supported |
| H5 | PC → DS | 0.29*** | 0.04 | 7.25 | Supported |
| H6 | PC → FR | 0.33*** | 0.04 | 8.25 | Supported |
| H7 | PR → DS | 0.18** | 0.06 | 3.00 | Supported |
| H8 | PR → FR | 0.16** | 0.06 | 2.67 | Supported |

Note. DIS = disconfirmation; PC = perceived complexity; PR = perceived risk; DS = dissatisfaction; FR = frustration; DI = discontinuance intention. $\beta$ = standardised path coefficient; SE = standard error.

**Figure 2. Structural Model Results**
**Alt text:** Structural model results for AI-IDLE discontinuance. Disconfirmation positively predicts dissatisfaction and frustration. Perceived complexity positively predicts dissatisfaction and frustration. Perceived risk also positively predicts dissatisfaction and frustration. Dissatisfaction and frustration both positively predict discontinuance intention, with frustration showing the strongest direct effect.

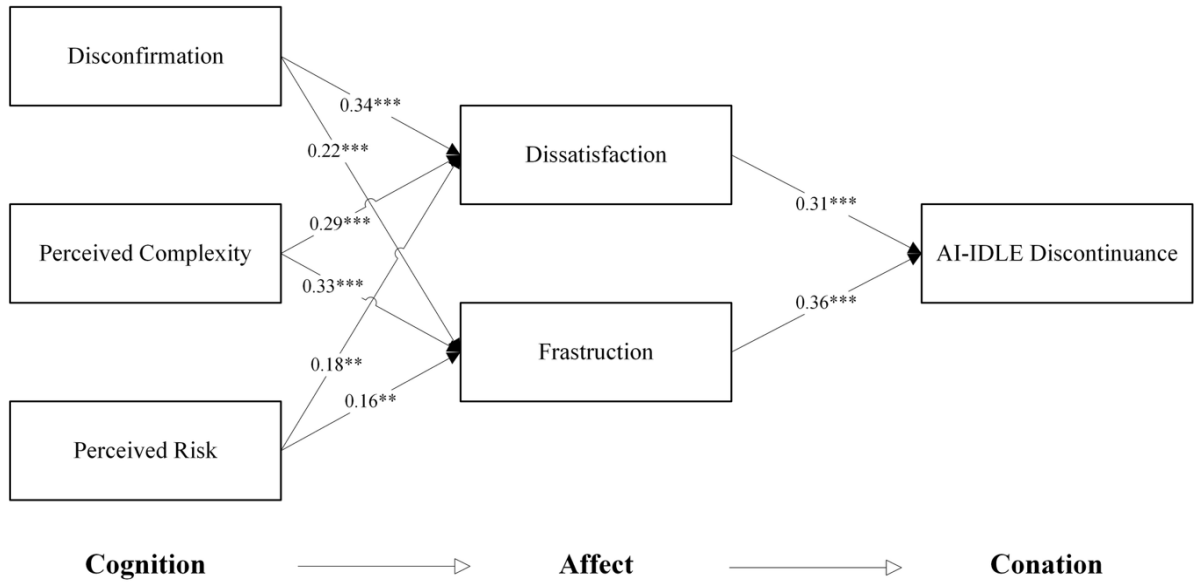

Note. *** $p < 0.001$; ** $p < 0.01$.

The mediation analysis results are reported in Table 9. Bootstrapping showed that dissatisfaction and frustration significantly mediated the relationships between the cognitive antecedents and discontinuance intention. Significant indirect effects were found for disconfirmation through dissatisfaction ($\beta = 0.11$, 95% CI [0.06, 0.17]) and frustration ($\beta = 0.08$, 95% CI [0.04, 0.13]), perceived complexity through dissatisfaction ($\beta = 0.09$, 95% CI [0.05, 0.14]) and frustration ($\beta = 0.12$, 95% CI [0.07, 0.18]), and perceived risk through dissatisfaction ($\beta = 0.06$, 95% CI [0.02, 0.11]) and frustration ($\beta = 0.06$, 95% CI [0.02, 0.10]). Since none of the confidence intervals included zero, all indirect effects were significant. Overall, the SEM results indicate that learners' negative cognitive evaluations of AI-IDLE contribute to discontinuance intention through dissatisfaction and frustration, with frustration showing the strongest direct effect on discontinuance intention.

**Table 9. Mediation Analysis Results (Bootstrapping)**

| Indirect Path | Indirect Effect ($\beta$) | Bootstrapped *SE* | 95% CI | Result |
| --- | --- | --- | --- | --- |
| DIS → DS → DI | 0.11 | 0.03 | [0.06, 0.17] | Significant |
| DIS → FR → DI | 0.08 | 0.02 | [0.04, 0.13] | Significant |
| PC → DS → DI | 0.09 | 0.02 | [0.05, 0.14] | Significant |
| PC → FR → DI | 0.12 | 0.03 | [0.07, 0.18] | Significant |
| PR → DS → DI | 0.06 | 0.02 | [0.02, 0.11] | Significant |
| PR → FR → DI | 0.06 | 0.02 | [0.02, 0.10] | Significant |

Note. DIS = disconfirmation; PC = perceived complexity; PR = perceived risk; DS = dissatisfaction; FR = frustration; DI = discontinuance intention. $\beta$ = standardised indirect effect; *SE* = standard error; CI = confidence interval.

### 5.2 Fuzzy-Set Qualitative Comparative Analysis (fsQCA)

Fuzzy-set qualitative comparative analysis (fsQCA) was conducted to complement the SEM results by identifying configurations of conditions associated with AI-IDLE discontinuance. Before the configurational analysis, all variables were calibrated into fuzzy-set scores. Following common fsQCA practice, three qualitative anchors were used: the 95th percentile for full membership, the 50th percentile for the crossover point, and the 5th percentile for full non-membership. The calibrated conditions included disconfirmation, perceived complexity, perceived risk, dissatisfaction, and frustration, while the outcome was AI-IDLE discontinuance. A necessity analysis was first performed to determine whether any single condition was necessary for the outcome. No condition reached the recommended consistency threshold of 0.90, indicating that AI-IDLE discontinuance could not be explained by any single condition alone.

A truth table was then constructed using the calibrated conditions. A frequency threshold was applied to ensure sufficient empirical representation, and a consistency threshold of 0.80 was used to identify configurations reliably associated with AI-IDLE discontinuance. As shown in Table 10, four sufficient configuration paths were identified. Path 1 reflects a disconfirmation–dissatisfaction route, Path 2 reflects a complexity–frustration route, Path 3 reflects a risk–dissatisfaction route, and Path 4 shows that perceived complexity and frustration can contribute to AI-IDLE discontinuance even when disconfirmation is absent. The overall solution coverage was 0.67, and the overall solution consistency was 0.87, indicating that the identified configurations explained a substantial proportion of AI-IDLE discontinuance with acceptable consistency. Overall, the fsQCA results suggest that AI-IDLE discontinuance is characterised by causal complexity and equifinality, as different combinations of cognitive and affective conditions can lead to the same outcome.

**Table 10. Configuration Paths of AI-IDLE Discontinuance**

| Condition | Path 1 | Path 2 | Path 3 | Path 4 |
|---|---|---|---|---|
| Disconfirmation | ● | ○ | | ⊗ |
| Perceived complexity | | ● | ○ | ● |
| Perceived risk | ⊖ | | ● | ○ |
| Dissatisfaction | ● | | ● | ○ |
| Frustration | ○ | ● | | ● |
| Raw coverage | 0.39 | 0.34 | 0.29 | 0.24 |
| Unique coverage | 0.11 | 0.08 | 0.06 | 0.04 |
| Consistency | 0.91 | 0.88 | 0.86 | 0.84 |
| Overall solution coverage | 0.67 | | | |
| Overall solution consistency | 0.87 | | | |

Note. ● indicates the presence of a core condition; ○ indicates the presence of a peripheral condition; ⊗ indicates the absence of a core condition; ⊖ indicates the absence of a peripheral condition; blank cells indicate "don't care" conditions.

## 6. Discussion
### 6.1 The Impact of Dissatisfaction and Frustration
The SEM results confirm that dissatisfaction and frustration are significant affective predictors of learners' discontinuance intention in AI-mediated informal digital learning of English (AI-IDLE). Dissatisfaction had a positive effect on discontinuance intention, indicating that learners who evaluate their AI-IDLE experience negatively are more likely to reduce or abandon their use of AI tools. This finding is consistent with post-adoption and expectation-confirmation perspectives, which suggest that users' continued engagement depends on whether actual experience meets prior expectations and produces sufficient perceived value (Lin et al., 2020; Zhou & Wang, 2025). In the AI-IDLE context, dissatisfaction may arise when AI tools fail to provide accurate explanations, pedagogically useful feedback, meaningful interaction, or reliable support for learners' self-directed English learning. Given that AI-IDLE is voluntary and weakly regulated by formal instructional structures, learners have considerable freedom to withdraw when the experience is no longer perceived as beneficial. This result therefore extends prior IDLE and AI-IDLE research by showing that negative affective evaluation is not merely an outcome of poor learning experience but a direct driver of discontinuance intention (Liu, Darvin, & Ma, 2025; Zhang & Tong, 2025).

Frustration showed an even stronger positive effect on discontinuance intention than dissatisfaction, suggesting that immediate emotional strain during AI-IDLE may be especially influential in learners' decisions to disengage. This finding aligns with research on technology-rich learning environments, where frustration is associated with obstacles, interruptions, usability problems, and difficulties in achieving learning goals (Novak et al., 2022, 2023). In AI-IDLE, frustration may occur when learners repeatedly receive irrelevant or inaccurate AI-generated responses, need to revise prompts excessively, encounter technical instability, or struggle to judge the reliability of AI output. Unlike dissatisfaction, which reflects a broader evaluative judgement of the learning experience, frustration captures the more situational emotional burden produced by impeded interaction. The stronger role of frustration implies that learners' discontinuance intention may be triggered not only by their overall negative assessment of AI-IDLE but also by repeated moments of interactional failure that make AI-supported learning feel effortful, uncertain, or cognitively taxing. This supports the cognition–affect–conation logic of the present study, demonstrating that affective responses translate learners' negative AI-IDLE experiences into behavioural withdrawal (Soyoof et al., 2023; Zhou & Zhang, 2025).

### 6.2 The Impact of Disconfirmation, Perceived Complexity, and Perceived Risk

The SEM results indicate that disconfirmation significantly increased both dissatisfaction and frustration, confirming its role as a key cognitive antecedent of negative affective responses in AI-IDLE. This suggests that when learners' actual experiences with AI tools fall short of their prior expectations, they are more likely to evaluate the learning experience negatively and experience emotional strain. This finding is consistent with expectation-disconfirmation theory, which argues that users form satisfaction judgements by comparing actual performance with expected performance (Carraher-Wolverton & Hirschheim, 2023; Lin et al., 2020). In AI-IDLE, learners may expect AI tools to provide accurate explanations, immediate feedback, natural interaction, and personalised support for informal English learning. When these expectations are not fulfilled, learners may perceive AI-IDLE as less useful or dependable than anticipated, thereby becoming dissatisfied and frustrated. This result extends recent AI-IDLE research by showing that learners' post-adoption affective responses depend not only on whether AI tools are available and innovative, but also on whether their actual learning value corresponds to learners' expectations (Guan et al., 2025; Zadorozhnyy & Lee, 2025).

Perceived complexity also had significant positive effects on dissatisfaction and frustration, indicating that learners who regard AI-IDLE as difficult to understand, manage, or integrate into their learning routines are more likely to develop negative affective responses. Although generative AI tools are often presented as accessible and efficient learning resources, effective AI-IDLE still requires learners to formulate appropriate prompts, interpret AI-generated feedback, verify linguistic accuracy, and decide how to incorporate AI responses into their self-directed learning practices. These demands may increase perceived effort and reduce the sense that AI tools offer convenient or reliable support. This finding aligns with prior research showing that complexity can inhibit technology adoption and use, particularly when users perceive a system as cognitively demanding or difficult to operate effectively (Hmoud et al., 2023; Liu, Darvin, & Ma, 2025). In the AI-IDLE context, complexity may be especially consequential because learners often use AI tools independently, without teacher scaffolding or institutional guidance. As a result, perceived complexity may undermine both the evaluative and emotional dimensions of learners' AI-IDLE experience.

Perceived risk had weaker but still significant positive effects on dissatisfaction and frustration, suggesting that learners' concerns about the possible negative consequences of AI-IDLE contribute to adverse affective responses, although less strongly than disconfirmation and complexity. These risks may include inaccurate AI-generated output, privacy concerns, over-reliance on AI, academic integrity issues, and reduced learner autonomy. This finding is consistent with research showing that perceived risk can weaken users' confidence in AI-assisted learning environments and increase negative attitudes towards AI use (Li, 2025; Wu et al., 2022; Zhang & Tong, 2025). In AI-IDLE, perceived risk may be less immediately disruptive than complexity or unmet expectations, but it can still erode learners' trust in AI tools as legitimate and dependable learning mediators. The significant effects of perceived risk therefore indicate that AI-IDLE discontinuance is shaped not only by usability and performance-related concerns, but also by broader judgements about the safety, reliability, and pedagogical appropriateness of AI-supported informal learning. Collectively, these findings support the cognition–affect–conation framework by demonstrating that disconfirmation, perceived complexity, and perceived risk operate as cognitive appraisals that trigger dissatisfaction and frustration, which subsequently contribute to discontinuance intention (Zhou & Zhang, 2025).

### 6.3 Configurational Mechanisms of AI-IDLE Discontinuance

The fsQCA results complement the SEM findings by showing that AI-IDLE discontinuance is not produced by a single isolated factor but by multiple configurations of cognitive and affective conditions. The necessity analysis showed that no individual condition reached the consistency threshold for necessity, indicating that disconfirmation, perceived complexity, perceived risk, dissatisfaction, and frustration are not individually indispensable for high discontinuance intention. Instead, the truth-table analysis identified four sufficient pathways, supporting the principle of equifinality in learners' AI-IDLE discontinuance. Path 1 represents a disconfirmation–dissatisfaction route, suggesting that when learners' actual AI-IDLE experiences fall short of expectations and produce dissatisfaction, they are likely to consider discontinuance even when perceived risk is not salient. This pathway is consistent with expectation-disconfirmation theory, which emphasises that unmet expectations can undermine satisfaction and weaken post-adoption continuance (Carraher-Wolverton & Hirschheim, 2023; Lin et al., 2020). Path 3 similarly highlights a risk–dissatisfaction route, indicating that perceived risk can lead to discontinuance when it is accompanied by dissatisfaction. This suggests that concerns about inaccurate output, privacy, over-reliance, academic integrity, or reduced autonomy become more consequential when they translate into a negative overall evaluation of AI-IDLE (Li, 2025; Wu et al., 2022; Zhang & Tong, 2025).

The remaining configurations further demonstrate that frustration constitutes an important affective route to AI-IDLE discontinuance. Path 2 reflects a complexity–frustration route, in which perceived complexity and frustration jointly contribute to discontinuance intention. This indicates that when learners experience AI-IDLE as difficult to manage, interpret, or integrate into their informal learning practices, the resulting frustration may prompt them to reduce or abandon use. Path 4 is especially informative because it shows that perceived complexity and frustration can still produce discontinuance even when disconfirmation is absent. In other words, learners may not necessarily feel that AI tools have failed to meet their expectations, but they may still discontinue use if the interaction process is experienced as burdensome, effortful, or emotionally taxing. This finding is consistent with research on frustration in technology-rich learning environments, where repeated obstacles and interactional difficulties can weaken engagement even when the technology is perceived as potentially useful (Novak et al., 2022, 2023). Overall, the fsQCA findings extend the SEM results by demonstrating causal asymmetry and configurational complexity: AI-IDLE discontinuance can emerge through different combinations of unmet expectations, perceived complexity, perceived risk, dissatisfaction, and frustration rather than through a uniform linear process (Schneider & Wagemann, 2012; Zhou & Zhang, 2025).

### 6.4 Theoretical and Practical Implications

This study contributes to AI-IDLE research by shifting attention from adoption and positive learning outcomes to discontinuance. The findings show that learners' use of AI tools for informal English learning is not automatically sustained after initial adoption. Instead, discontinuance is shaped by negative cognitive appraisals and affective responses. By applying the cognition–affect–conation framework, this study clarifies how disconfirmation, perceived complexity, and perceived risk generate dissatisfaction and frustration, which then increase discontinuance intention. The combined SEM and fsQCA results further show that AI-IDLE discontinuance is both linear and configurational, as different combinations of cognitive and affective conditions can lead to the same behavioural outcome.

Practically, the findings suggest that educators, developers, and institutions should reduce the conditions that make AI-IDLE frustrating or unsatisfactory. Educators can support learners by teaching effective prompting, critical evaluation of AI output, and responsible integration of AI feedback into English learning. Developers should improve the accuracy, transparency, usability, and pedagogical relevance of AI tools. Institutions should provide guidance on AI literacy, privacy, academic integrity, and responsible use. These measures may reduce perceived complexity and risk, manage learners' expectations, and support more sustainable engagement with AI-mediated informal English learning.

**6.5 Limitations and Future Directions**
This study has several limitations. First, the data were collected through a cross-sectional survey, which limits the ability to make strong causal claims about the relationships among cognitive appraisals, affective responses, and discontinuance intention (Y. Du, Tang, et al., 2026; Y. Du, Yuan, et al., 2026; Tang, Jia, et al., 2026; C. Wang et al., 2026; W. Zhang et al., 2026). Future research could use longitudinal or experimental designs to examine how learners' AI-IDLE experiences and discontinuance behaviours change over time (Y. Du, Li, et al., 2026; Y. Du & He, 2026c, 2026b, 2026a; Tang, Lau, et al., 2026). Second, the study measured discontinuance intention rather than actual discontinuance behaviour (C. Du et al., 2025; Y. Du, 2025b, 2025a, 2026; Y. Du et al., 2025). Although intention is an important predictor of behaviour, future studies could collect behavioural data, usage logs, or follow-up reports to examine whether learners actually reduce, suspend, or abandon AI tools (Chen et al., 2022; Y. Du, 2023, 2024; Y. Du et al., 2024; He & Du, 2024; C. Wang et al., 2024; Zou et al., 2023, 2024).

Third, the sample consisted of Chinese university students with prior AI-IDLE experience, so the findings may not fully generalise to learners in other educational, cultural, or linguistic contexts. Future research could compare learners across countries, age groups, proficiency levels, and learning environments. Finally, this study focused on disconfirmation, perceived complexity, perceived risk, dissatisfaction, and frustration. Future studies could include additional factors, such as AI literacy, trust, perceived usefulness, learner autonomy, task type, teacher guidance, and platform characteristics, to provide a more comprehensive understanding of AI-IDLE discontinuance.

**7. Conclusion**
This study examined university students' discontinuance intention towards AI-mediated informal digital learning of English through the cognition–affect–conation framework. The SEM results showed that dissatisfaction and frustration directly increased discontinuance intention, while disconfirmation, perceived complexity, and perceived risk indirectly contributed to discontinuance by increasing these negative affective responses. Among the affective factors, frustration showed the strongest direct effect, suggesting that repeated difficulty or emotional strain during AI-IDLE may be especially important in explaining why learners reduce or abandon AI tools. The fsQCA results further showed that AI-IDLE discontinuance is configurational rather than driven by a single necessary condition. Different combinations of disconfirmation, perceived complexity, perceived risk, dissatisfaction, and frustration can lead to high discontinuance intention. Overall, the findings suggest that sustaining AI-IDLE requires more than providing access to AI tools. Learners also need manageable, trustworthy, and pedagogically useful AI-supported experiences. Reducing complexity, risk, dissatisfaction, and frustration is therefore essential for supporting more sustainable engagement with AI-mediated informal English learning.

**References**
Carraher-Wolverton, C., & Hirschheim, R. (2023). Utilizing expectation disconfirmation theory to develop a higher-order model of outsourcing success factors. *Journal of Systems and Information Technology*, *25*(1), 1–29. https://doi.org/10.1108/JSIT-05-2022-0133
Chen, X., Du, Y., Qu, M., & Gao, S. (2022). *A study on the effect of L1 to L2 transfer on the production of idiomatic expressions in L2 among mandarin-speaking intermediate learners of English:* 2021 International Conference on Public Art and Human Development ( ICPAHD 2021). https://doi.org/10.2991/assehr.k.220110.117
Du, C., Tang, M., Wang, C., Zou, B., Xia, Y., & Du, Y. (2025). Who is most likely to accept AI chatbots? A sequential explanatory mixed-methods study of personality and ChatGPT acceptance for language learning. *Innovation in Language Learning and Teaching*, 1–22. https://doi.org/10.1080/17501229.2025.2555515
Du, Y. (2023). A corpus-based study to evaluate the generativist explanation of children's error patterns in questions. *Journal of Language Teaching*, *3*(3), 26–38. https://doi.org/10.54475/jlt.2023.007

Du, Y. (2024). A streamlined approach to scale adaptation: Enhancing validity and feasibility in educational measurement. *Journal of Language Teaching*, *4*(3), 18–22. https://doi.org/10.54475/jlt.2024.017

Du, Y. (2025a). *Confirmation bias in generative AI chatbots: Mechanisms, risks, mitigation strategies, and future research directions* (Version 1). arXiv. https://doi.org/10.48550/ARXIV.2504.09343

Du, Y. (2025b). The impact of emojis on verbal irony comprehension in computer-mediated communication: A cross-cultural study. *International Journal of Human–Computer Interaction*, *41*(8), 4979–4986. https://doi.org/10.1080/10447318.2024.2356398

Du, Y. (2026). *Examining users' behavioural intention to use OpenClaw through the cognition-affect-conation framework* (Version 2). arXiv. https://doi.org/10.48550/ARXIV.2603.11455

Du, Y., & He, H. (2026a). *Enabling and inhibitory pathways of students' AI use concealment intention in higher education: Evidence from SEM and fsQCA* (Version 1). arXiv. https://doi.org/10.48550/ARXIV.2604.10978

Du, Y., & He, H. (2026b). *Enabling and inhibitory pathways of university students' willingness to disclose AI use: A cognition-affect-conation perspective* (Version 1). arXiv. https://doi.org/10.48550/ARXIV.2604.21733

Du, Y., & He, H. (2026c). *Examining EAP students' AI disclosure intention: A cognition-affect-conation perspective* (Version 1). arXiv. https://doi.org/10.48550/ARXIV.2604.10991

Du, Y., He, H., & Chu, Z. (2024). Cross-cultural nuances in sarcasm comprehension: A comparative study of Chinese and American perspectives. *Frontiers in Psychology*, *15*, 1349002. https://doi.org/10.3389/fpsyg.2024.1349002

Du, Y., Li, J., He, H., Wang, C., & Zou, B. (2026). *A sequential explanatory mixed-methods study on the acceptance of a social robot for EFL speaking practice among Chinese primary school students: Insights from the Computers Are Social Actors (CASA) paradigm* (Version 1). arXiv. https://doi.org/10.48550/ARXIV.2604.12789

Du, Y., Tang, M., Jia, K., Wang, C., & Zou, B. (2026). Are teachers addicted to AI? Analysing factors influencing dependence on generative AI through the I-PACE model. *Journal of Computer Assisted Learning*, *42*(1), e70174. https://doi.org/10.1002/jcal.70174

Du, Y., Wang, C., Zou, B., & Xia, Y. (2025). Personalizing AI tools for second language speaking: The role of gender and autistic traits. *Frontiers in Psychiatry*, *15*, 1464575. https://doi.org/10.3389/fpsyt.2024.1464575

Du, Y., Yuan, Y., Wang, C., He, H., & Jia, K. (2026). Was this person being ironic? The role of emojis in irony comprehension and memory in computer-mediated communication: insights from the UK and China. *Telematics and Informatics*, *106*, 102390. https://doi.org/10.1016/j.tele.2026.102390

Guan, L., Zhang, E. Y., & Gu, M. M. (2025). Examining generative AI–mediated informal digital learning of English practices with social cognitive theory: A mixed-methods study. *ReCALL*, *37*(3), 315–331. https://doi.org/10.1017/S0958344024000259

Guangxiang Leon Liu, Zou, M. M., Soyoof, A., & Chiu, M. M. (2025). Untangling the relationship between AI-mediated informal digital learning of English (AI-IDLE), foreign language enjoyment and the ideal L2 self: Evidence from Chinese university EFL students. *European Journal of Education*, *60*(1), e12846. https://doi.org/10.1111/ejed.12846

He, H., & Du, Y. (2024). The effectiveness of dialogical argumentation in supporting low-level EAP learners' evidence-based writing: A longitudinal study. In B. Zou & T. Mahy (Eds), *English for Academic Purposes in the EMI Context in Asia: XJTLU Impact* (pp. 45–75). Springer Nature Switzerland. https://doi.org/10.1007/978-3-031-63638-7_3

Hmoud, H., Al-Adwan, A. S., Horani, O., Yaseen, H., & Zoubi, J. Z. A. (2023). Factors influencing business intelligence adoption by higher education institutions. *Journal of Open Innovation: Technology, Market, and Complexity*, *9*(3), 100111. https://doi.org/10.1016/j.joitmc.2023.100111

Kline, R. B. (2023). *Principles and practice of structural equation modeling* (Fifth edition). The Guilford Press.

Klotz, A. C., Swider, B. W., & Kwon, S. H. (2023). Back-translation practices in organizational research: Avoiding loss in translation. *Journal of Applied Psychology*, *108*(5), 699–727. https://doi.org/10.1037/apl0001050

Lee, J. S., Xie, Q., & Lee, K. (2024). Informal digital learning of English and L2 willingness to communicate: Roles of emotions, gender, and educational stage. *Journal of Multilingual and Multicultural Development*, *45*(2), 596–612. https://doi.org/10.1080/01434632.2021.1918699

Lee, J. S., Yeung, N. M., & Osburn, M. B. (2024). Foreign language enjoyment as a mediator between informal digital learning of english and willingness to communicate: A sample of hong kong EFL secondary students. *Journal of Multilingual and Multicultural Development*, *45*(9), 3613–3631. https://doi.org/10.1080/01434632.2022.2112587

Li, W. (2025). A study on factors influencing designers' behavioral intention in using AI-generated content for assisted design: Perceived anxiety, perceived risk, and UTAUT. *International Journal of Human–Computer Interaction*, *41*(2), 1064–1077. https://doi.org/10.1080/10447318.2024.2310354

Lin, J., Lin, S., Turel, O., & Xu, F. (2020). The buffering effect of flow experience on the relationship between overload and social media users' discontinuance intentions. *Telematics and Informatics*, *49*, 101374. https://doi.org/10.1016/j.tele.2020.101374

Liu, G., Darvin, R., & Ma, C. (2025). Exploring AI-mediated informal digital learning of English (AI-IDLE): A mixed-method investigation of Chinese EFL learners' AI adoption and experiences. *Computer Assisted Language Learning*, *38*(7), 1632–1660. https://doi.org/10.1080/09588221.2024.2310288

Liu, G., Ma, C., Bao, J., & Liu, Z. (2025). Toward a model of informal digital learning of English and intercultural competence: A large-scale structural equation modeling approach. *Computer Assisted Language Learning*, *38*(3), 342–368. https://doi.org/10.1080/09588221.2023.2191652

Marikyan, D., Papagiannidis, S., & Alamanos, E. (2023). Cognitive dissonance in technology adoption: A study of smart home users. *Information Systems Frontiers*, *25*(3), 1101–1123. https://doi.org/10.1007/s10796-020-10042-3

Novak, E., McDaniel, K., Daday, J., & Soyturk, I. (2022). Frustration in technology-rich learning environments: A scale for assessing student frustration with e-textbooks. *British Journal of Educational Technology*, *53*(2), 408–431. https://doi.org/10.1111/bjet.13172

Novak, E., McDaniel, K., & Li, J. (2023). Factors that impact student frustration in digital learning environments. *Computers and Education Open*, *5*, 100153. https://doi.org/10.1016/j.caeo.2023.100153

Podsakoff, P. M., Podsakoff, N. P., Williams, L. J., Huang, C., & Yang, J. (2024). Common method Bias: It's bad, it's complex, it's widespread, and it's not easy to fix. *Annual Review of Organizational Psychology and Organizational Behavior*, *11*(1), 17–61. https://doi.org/10.1146/annurev-orgpsych-110721-040030

Ranjan, A., & Mukherjee, S. (2026). Unveiling the dark side of technology adoption: Discontinuance intention through negative expectation and cognitive dissonance theories. *Journal of Marketing Theory and Practice*, 1–20. https://doi.org/10.1080/10696679.2026.2653978

Schneider, C. Q., & Wagemann, C. (2012). *Set-theoretic methods for the social sciences: A guide to qualitative comparative analysis* (1st edn). Cambridge University Press. https://doi.org/10.1017/CBO9781139004244

Song, C., & Zhou, S. (2026). Push or Pull? Understanding switching intentions from human services to GAI agents through the PPM framework. *Computers in Human Behavior*, *176*, 108874. https://doi.org/10.1016/j.chb.2025.108874

Soyoof, A., Reynolds, B. L., Vazquez-Calvo, B., & McLay, K. (2023). Informal digital learning of English (IDLE): A scoping review of what has been done and a look towards what is to come. *Computer Assisted Language Learning*, *36*(4), 608–640. https://doi.org/10.1080/09588221.2021.1936562

Tang, M., Jia, K., He, H., Wang, C., Zou, B., & Du, Y. (2026). Acceptance and engagement in artificial intelligence–supported reading among primary school learners of english as a foreign language. *International Journal of Applied Linguistics*, ijal.70204. https://doi.org/10.1111/ijal.70204

Tang, M., Lau, K.-L., & Du, Y. (2026). Effects and moderators of dialogic reading on children's reading literacy: A three-level meta-analysis on studies from 2000 to 2025. *International Journal of Educational Research*, *137*, 102963. https://doi.org/10.1016/j.ijer.2026.102963

Wang, C., Du, Y., & Zou, B. (2026). Learners' acceptance and use of multimodal artificial intelligence (AI)-generated content in AI-mediated informal digital learning of English. *International Journal of Applied Linguistics*, *36*(1), 927–940. https://doi.org/10.1111/ijal.12827

Wang, C., Zou, B., Du, Y., & Wang, Z. (2024). The impact of different conversational generative AI chatbots on EFL learners: An analysis of willingness to communicate, foreign language speaking anxiety, and self-perceived communicative competence. *System*, *127*, 103533. https://doi.org/10.1016/j.system.2024.103533

Wang, X., Gao, Y., Reynolds, B. L., & Wang, Q. (2025). Exploring Chinese EFL learners' beliefs about AI-mediated informal digital learning of English: Insights from Q Methodology. *Porta Linguarum Revista Interuniversitaria de Didáctica de Las Lenguas Extranjeras*, (XIII), 131–146. https://doi.org/10.30827/portalin.viXIII.31925

Ward, M. K., & Meade, A. W. (2023). Dealing with careless responding in survey data: Prevention, identification, and recommended best practices. *Annual Review of Psychology*, *74*(1), 577–596. https://doi.org/10.1146/annurev-psych-040422-045007

Wu, R. (2023). The relationship between online learning self-efficacy, informal digital learning of English, and student engagement in online classes: The mediating role of social presence. *Frontiers in Psychology*, *14*, 1266009. https://doi.org/10.3389/fpsyg.2023.1266009

Wu, W., Zhang, B., Li, S., & Liu, H. (2022). Exploring factors of the willingness to accept AI-assisted learning environments: An empirical investigation based on the UTAUT model and perceived risk theory. *Frontiers in Psychology*, *13*, 870777. https://doi.org/10.3389/fpsyg.2022.870777

Xia, M., & Guo, S. (2025). Understanding learners' perceptions of artificial intelligence-mediated Informal Digital Learning of English: A Q methodology approach. *Acta Psychologica*, *261*, 105980. https://doi.org/10.1016/j.actpsy.2025.105980

Yang, S., Xu, W., Liu, R., & Yu, Z. (2025). Influencing and moderating variables in informal digital learning of English through a structural equation model. *Computer Assisted Language Learning*, *38*(7), 1452–1487. https://doi.org/10.1080/09588221.2023.2280645

Zadorozhnyy, A., & Lee, J. S. (2025). Informal Digital Learning of English and willingness to communicate in a second language: Self-efficacy beliefs as a mediator. *Computer Assisted Language Learning*, *38*(4), 669–689. https://doi.org/10.1080/09588221.2023.2215279

Zhang, T., & Tong, Q. (2025). The technostress of ChatGPT usage: How do perceived AI characteristics affect user discontinuous usage through AI anxiety and user negative attitudes? *International Journal of Human–Computer Interaction*, *41*(16), 9918–9929. https://doi.org/10.1080/10447318.2024.2429889

Zhang, W., Zou, B., & Du, Y. (2026). Teachers' perceptions of the current practices and challenges in English for academic purposes: A survey study at universities in Shanghai, China. *International Journal of English for Academic Purposes: Research and Practice*, *6*(1), 7–28. https://doi.org/10.3828/ijeap.2026.2

Zhang, Y., & Liu, G. (2024). Revisiting informal digital learning of English (IDLE): A structural equation modeling approach in a university EFL context. *Computer Assisted Language Learning*, *37*(7), 1904–1936. https://doi.org/10.1080/09588221.2022.2134424

Zhang, Y., & Liu, G. L. (2025). Examining the impacts of learner backgrounds, proficiency level, and the use of digital devices on informal digital learning of English: An explanatory mixed-method study. *Computer Assisted Language Learning*, *38*(5–6), 1113–1140. https://doi.org/10.1080/09588221.2023.2267627

Zhou, T., & Wang, M. (2025). Examining generative AI user discontinuance from a dual perspective of enablers and inhibitors. *International Journal of Human–Computer Interaction*, *41*(20), 13140–13150. https://doi.org/10.1080/10447318.2025.2470280

Zhou, T., & Zhang, C. (2025). Examining generative AI user intermittent discontinuance from a C-A-C perspective. *International Journal of Human–Computer Interaction*, *41*(10), 6377–6387. https://doi.org/10.1080/10447318.2024.2376370

Zou, B., Du, Y., Wang, Z., Chen, J., & Zhang, W. (2023). An investigation into artificial intelligence speech evaluation programs with automatic feedback for developing EFL learners' speaking skills. *Sage Open*, *2023*(7). https://doi.org/10.1177/21582440231193818

Zou, B., Liviero, S., Ma, Q., Zhang, W., Du, Y., & Xing, P. (2024). Exploring EFL learners' perceived promise and limitations of using an artificial intelligence speech evaluation system for speaking practice. *System*, *126*, 103497. https://doi.org/10.1016/j.system.2024.103497